\documentstyle[preprint,prl,aps]{revtex}
\begin{document}
\draft
\title{Coulomb Drag of Edge Excitations in the Chern--Simons Theory
 of the Fractional Quantum Hall Effect}

\author{Dror Orgad and Shimon Levit}

\address{Department of Condensed Matter, The Weizmann Institute of Science,\\
Rehovot 76100, Israel}

\date{\today}

\maketitle

\begin{abstract}

Long range Coulomb interaction between the edges of a Hall bar changes the
nature  of the gapless edge excitations.  Instead of independent modes 
propagating in opposite  directions on each edge  as expected for a short
range interaction one finds elementary excitations living simultaneously 
on both edges, i.e.  composed of 
correlated density waves propagating in the {\em same direction} on
opposite edges.  	
We discuss the microscopic features of this Coulomb drag of excitations   in the fractional
quantum  Hall regime within the framework of the bosonic Chern-Simons Landau-Ginzburg 
theory. The dispersion law of these novel excitations  is non linear and 
depends on  the distance between the edges as well  as  on the current that flows through the sample. 
The latter dependence indicates a possibility of parametric excitation of these modes. 
The bulk distributions of the density and currents 
of the edge excitations differ significantly for short and
long range  interactions.

\end{abstract}


The Integer and Fractional Quantum Hall Effects occur in a two dimensional electron gas placed in a 
strong perpendicular magnetic field. Interesting phenomena can take place at the edges of this gas. 
The simplest is related to the edge currents which are instrumental in our understanding of the 
transport properties of the QHE \cite{Halperin82,Buttiker88,Beenakker90,Chklovskii92}. 
The character of the excitations in the Quantum Hall system is 
also effected by the presence of the edges. While the bulk excitations have a finite gap the 
excitations which are localized near the edges are found to be gapless. They are chiral, that is 
moving only in one direction along the edge, and can be described by the chiral Luttinger model 
\cite{Wen92} provided one disregards their coupling to other modes. Recently the possibility to 
probe the structure of the fractional Hall states by studying the tunneling between edges 
attracted much attention \cite{Chamon93,Kane94}.  

In this paper we study the edge excitations within the framework of a bosonic Chern--Simons Landau--
Ginsburg (CSLG) theory. This theory \cite{Zhang89} was proposed as a useful supplement and extension 
of Laughlin's fully microscopic many body  theory of the FQHE.  Recently, the edge effects 
were studied using this theory \cite{Nagaosa94} 
and Wen's results were rederived. Our goal is to investigate the effects of the inter-edge 
interactions. In \cite{Oreg95} novel effects were predicted due to this interaction which are related 
to the redefinition of the gapless modes into a Bogolubov  transformed combination of modes living on both 
edges and propagating in the same direction. Our analysis goes one step further and represents the 
microscopic picture of this phenomenon. We use the realistic Coulomb interaction between the edges 
and derive the modified modes and their density and current distributions. These clearly indicate  
that one deals with the Coulomb drag of excitations on one edge by the other. We derive a non linear 
dispersion relation for the modified excitation modes and find that it depends on the width of the 
Hall bar and on the total current which flows through it. We point out that the latter dependence opens 
a possibility of a parametric excitation of the novel modes.          

The starting point of our analysis is the mean field equations derived by 
minimizing the CSLG action for the fractional quantum 
Hall effect\cite{Zhang89}.  It will be convenient to write these  
 equations in a hydrodynamic form using 
the polar decomposition of the bosonic field $\phi=\sqrt{\rho}e^{i \theta}$ 
and introducing the velocity fields 
$v_i={\hbar\over{m}}\partial_i\theta+{e\over {mc}}(a_i+A_i)$  with the vector 
potential $\bf{A}$  taken such as to produce a constant magnetic field along the 
z direction,
\begin{mathletters}
\begin{eqnarray}
\label{bfield}
&&\epsilon_{ij}\partial_i v_j=1-\rho\\
\label{efield}
&&\partial_t\partial_i\theta-\partial_t v_i-\partial_i a_0=\epsilon_{ij}\rho 
v_j\\
\label{cont}
&&\partial_t\rho=-\partial_i(\rho v_i)\\
\label{sch}
&&\partial_t\theta=-{1\over 2}v_i^2+{1\over 2}{\partial_i^2\sqrt{\rho}\over 
\sqrt{\rho}}+a_0-V(x)-\rho-{\nu\over{2\pi}}\int  U({\bf r-r
'})\rho({\bf r}')d^2{\bf r}'\; . 
\end{eqnarray}
\end{mathletters}
Here and throughout the paper length is measured in units of the magnetic 
length $l=\sqrt{\hbar c/eB}$, time in units of inverse cyclotron frequency 
$\omega_c=eB/mc$, energy is normalized by $\hbar\omega_c$ and the density by 
its bulk value $\bar\rho=\nu B/\phi_0$ where $\nu=1/(2n+1)$ is the filling
 factor. All the  quantities  apart of  $\theta$ and $a_0$  are gauge invariant and the
latter appear in the gauge invariant combination $\partial_t\theta - a_0$.
In order to correctly reproduce the energetics of the noninteracting limit we have followed
  Ref.\cite{Ezawa92}  and included  a $\delta$-function type repulsive force (the term
before the last in (\ref{sch}))  with a strength of 
$2\pi\hbar^2/\nu m$.  Our confining potential
is assumed to rise fast enough to avoid the occurrence  of 
an alternating sequence of compressible and incompressible strips along the edge
\cite{Beenakker90,Chklovskii92}.   Apart from this requirement the detailed nature of 
$V(x)$ is of no importance for our applications.

Our aim is to obtain the 
edge  excitations as the RPA modes  of the theory , i.e.  the eigenmodes of
the  above equations linearized around a static solution. We  first consider a 
  single edge.  The translation invariance in the $y $  direction suggests
looking for a static solution in the form:
\begin{equation}
\label{sol}
\theta=-x_0y-\mu(x_0)t \;\;,\;\; \rho = \rho(x) \;\;,\;\; 
v_x=0 \;\;,\;\; v_y =  v_y(x) \;\;, \;\;a_0 = a_0(x) \; .
\end{equation}
We assume for definiteness an infinitely  high wall situated at
$x\leq 0$  and accordingly set  the density  to zero  at 
the wall. The only gauge freedom which preserves the form of the above  solution is adding
arbitrary constants to $x_0$ and $\mu$.  We  fix this freedom by requiring that  the statistical
potentials $a_0$, $a_i$  vanish at $x = 0$ and choosing ${\bf A}=(0,Bx,0)$. This assures 
that solutions with different values of $x_0$ and $\mu$ are not  gauge  transforms of each other. 
 Formally $x_0$ is the conserved momentum along the edge and its physical effect on the
condensate $\phi$ is similar to the guiding center coordinate of Landau levels --
changing its  value  translates  $\phi$  in the x--direction.  For a fixed $x_0$  the value
of  $\mu$ is determined by requiring  that $\rho$ approaches    $\bar\rho$  
far from  the edge.  It  can be  shown   that $\mu$ is 
  the energy which is needed in order to add a particle to the edge.
Representative examples of the density and current density profiles for solutions
(\ref{sol}) in the  case of short range interactions $U=(2\pi\hbar^2\lambda_s/ \nu m)
\delta({\bf r-r'})$ are shown in Fig. \ref{fig1}.   One finds a {\em one
parameter family of static solutions} depending on $x_0$  and differing  by  the 
density of particles at the edge.  With our boundary conditions the velocity at the edge  
is $v_y(0)=-x_0$.  Using this and the fact that the velocity
falls to zero in the bulk  we  integrate Eq.~(\ref{bfield})  and find that 
$-x_0$ is the {\em excess charge} per  unit length along the edge relative to a step-like
constant density  profile. 

For a Coulomb interaction $U=2\pi\nu^{-1}\lambda_c\hbar\omega_c/\mid\bf r-r'\mid$ with 
$\lambda_c=\nu e^2/2\pi\epsilon l\,\hbar\omega_c$ and 
 a constant neutralizing 
background extending up  to the wall one obtains   distributions similar to 
Fig. \ref{fig1}.  The  essential difference between the two cases 
is found in the large $x$ behavior of the solutions. By a simple 
 analysis of Eqs.~(\ref{bfield}-\ref{sch}) one finds that for the short 
range interaction all the relevant quantities decay exponentially to their bulk 
values with a decay constant $\alpha=\sqrt{2(\lambda_s+1)-2\sqrt{(\lambda_s+1)^2-1}}$. For
 the Coulomb interaction the density and the velocity behave at 
large distances according to $1-2\lambda_c x_0/x^2$ and 
$-2\lambda_c x_0/x$ respectively.   

Even more pronounced difference is found in the dependence  of
$\mu$ on  $x_0$ which is displayed in Fig. \ref{fig2}. For the short range interaction
the  chemical potential tends to ${1 \over 2}(\lambda_s+1)$ for large positive values of $x_0$ 
and increases as the charges are pushed against the wall, i.e. for decreasing $x_0$. In the
case of  the Coulomb interaction, assuming that the length of the sample $L$ is very large compared 
to $x_0$, one finds that the leading contribution to $\mu$ is given by $-2\lambda_c x_0 \ln L$. 
This term corresponds to the electrostatic charging energy of the excess 
charge ($-x_0$) had it been  uniformly distributed over a strip of width $l$ along the edge. 
Its change of sign at $x_0=0$ reflects the  tendency of the 
system to remain neutral. The remaining part of the chemical 
potential $\tilde\mu=\mu+2\lambda_c x_0 \ln L$ is due to the kinetic energy and the deviation of 
the excess charge distribution from that of a uniform strip. As one can observe from Fig. \ref{fig2} 
the dependence of $\tilde\mu$ on $x_0$ varies considerably as the strength of the interaction is changed. 

Linearizing  Eqs.~(\ref{bfield}-\ref{sch}) around one of the 
solutions (\ref{sol}) we obtain the RPA equations 
\begin{mathletters}
\begin{eqnarray}
\label{lbfield}
&&\epsilon_{ij}\partial_i \delta v_j=-\delta\rho\\
\label{lefield}
&&\partial_t\partial_i\delta\theta-\partial_t\delta v_i-\partial_i\delta a_0=
\epsilon_{iy}v_y\delta\rho+\epsilon_{ij}\rho\delta v_j\\
\label{lcont}
&&\partial_t\delta\rho=-v_y\partial_y\delta\rho-\partial_i(\rho\delta v_i)\\
\label{lsch}
&&\partial_t\delta\theta=-v_y\delta v_y+f(\rho,\delta\rho)+{\partial_y^2\delta
\rho\over{4\rho}}+\delta a_0-\delta\rho-{\nu\over{2\pi}}\int  U({\bf r-r
'})\delta\rho({\bf r}')d^2{\bf r}'\; .
\end{eqnarray}
\end{mathletters}
where $f$ denotes  the part of the linearized  "quantum pressure" term, i.e. the
second term on the r.h.s. of  (\ref{sch}), containing $x$ derivatives. 

Our  main method of solving Eqs.~(\ref{lbfield}-\ref{lsch}) 
will be to  use the  continuous one parameter {\em family of static
solutions} described above in order  to find  the   {\em gapless} branch  of  
the RPA  eigenmodes. The derivative of the solution (\ref{sol}) with respect to
$x_0$ constitute a static solution of the RPA equations (\ref{lbfield}-\ref{lsch}). We denote 
it by $\delta\theta^{(s)}\:,\:
\delta\rho^{(s)}(x)\:,\:\delta v_y^{(s)}(x)\:,\:\delta a_0^{(s)}(x)$. 
Motivated by this observation and concentrating on the long wavelength limit we set
\begin{eqnarray}
\label{lsol}
\nonumber &&\delta\rho=\delta\rho^{(s)}(x)cos(ky-\omega t)\\
          &&\delta v_y=\delta v_y^{(s)}(x)cos(ky-\omega t)\\
\nonumber &&\delta a_0=\delta a_0^{(s)}(x)cos(ky-\omega t)\\
\nonumber &&\delta\theta=-{1\over k}sin(ky-\omega t) \; .
\end{eqnarray}
Clearly these density and velocity distributions are concentrated along the edge. 
With the expression (\ref{ldisp}) for $\omega$  given below also the gauge invariant 
combination $\partial_t\delta\theta - \delta a_0$   vanishes far away from the edge. 
Inserting these functions into Eq.~(\ref{lsch}) and neglecting the term proportional to 
$k^2$ we find that they indeed represent a solution of the equation provided we  choose
properly   the dispersion relation $\omega = \omega(k)$.  In the case of the short range
interaction we find  directly 
\begin{equation}
\label{ldisp}
\omega=-{l^2\over\hbar}{\partial\mu\over{\partial x_0}}k(1+kl) \; .
\end{equation}
where we have restored the units  of dimensions. In this expression we have included the 
second order term in $k$ which can be obtained 
after multiplying the first three fields in (\ref{lsol}) by $1+kl$. 
In the case of the Coulomb interaction  after  the  substitution of $\delta\rho$ 
the last term in Eq.~(\ref{lsch}) becomes 
$-2\lambda_c cos(ky-\omega t)\int_0^{\infty}dx'\delta \rho^{(s)} (x')
K_0(|k||x-x'|)$ where $K_0(x)$ is the modified 
Bessel function. Working  in the region $x\ll k^{-1}$ and assuming  
 that $k^{-1}$ is much larger then the width of the region where $\delta\rho^{(s)}$ 
is appreciable we can use the 
approximation $K_0(|k||x-x'|)\approx \ln(2e^{-\gamma}/
|k||x-x'|)$ where $\gamma$ is the Euler constant. As a result we 
find that Eq.~(\ref{lsch}) is satisfied to first order in $k$ when the 
dispersion law is modified by an extra logarithmic term, cf.,  
Refs\cite{Giovanazzi94,Volkov85},
\begin{equation}
\label{cdisp}
\omega=-{l^2\over\hbar}{\partial\tilde\mu\over{\partial x_0}}k+{\nu\over\pi}
{e^2\over{\epsilon\hbar}}k\,\ln\left({2e^{-\gamma}\over{|k|l}}\right)\; .
\end{equation}
Note that $\tilde\mu$ rather then $\mu$ (cf., Fig \ref{fig2}) enters this expression.

Integrating the linearized continuity equation (\ref{lcont}) we find that the 
above solution should be supplemented by a velocity field in the $x$ direction:
\begin{equation}
\label{vxfield}
\delta v_x={1\over\rho}(\omega\delta v_y^{(s)}-k\delta a_0^{(s)} +\omega)sin(ky-\omega t)
\end{equation}
which corresponds to a current $\rho \delta v_x$ vanishing  on the wall as well
as far away from it. One can also  check that the other two equations in the linearized set
are satisfied up to  first order in $k$.

For the short range interaction the linearity of the dispersion  relation and  the fact
 that $\partial\mu/\partial x_0\; < \;0$ imply the chirality of the waves.  The wave
velocity is determined by the compressibility of the edge $\kappa=(\bar\rho x_0^2\;
\partial\mu/\partial x_0)^{-1}$
which is controlled by the
wall potential.  It vanishes when the edge of the static solution is far away from the
wall (i.e. at large positive $x_0$)  for which case $\omega\sim k^3$  in
agreement with Ref.\cite{Giovanazzi94}.   
In the  Coulomb case the  
long wavelength limit  of the dispersion curve is dominated by the logarithmic part  so
that  although $\partial\tilde\mu/\partial x_0$   can have either sign  the excitations are
chiral as before. For distances which are larger than $k^{-1}$ the Bessel function $K_0$
decays exponentially  and we expect a crossover in the excitation profiles from the
Coulomb power law behavior to an exponential decay.

We go on now to consider the case of a wide and long Hall bar defined by hard walls at $x=\pm {W\over2}$. 
According to our strategy we first show that again a {\em  family}  of
static solutions of Eqs.~(\ref{bfield}-\ref{sch})  exists. This will now be a {\em two
parameter}  family  which we find  by appropriately gluing  together two solutions of the 
single wall case. Assuming that the two solutions correspond to the values of $x_0$ and
$\mu$ which are $(x_1,\mu_1)$  and $(x_2,\mu_2)$  respectively  and denoting the solutions 
by superscripts $1$ and $2$   we set 
\begin{eqnarray}
\label{bsol1}
\nonumber &&\theta=-x_1y-\mu_1t\;\; {\rm  (short\;\;
range)}\;\;\;,\;\;\;\theta=-x_1y-(\mu_1-2\lambda_c x_2 \ln {L\over W})t \;\;{\rm  (Coulomb)} \nonumber \\
 &&\rho=\rho^{(1)}(x+{W\over2})\Theta(-x)+
                 \rho^{(2)}({W\over2}-x)\Theta(x)  \nonumber\\
 &&v_x = 0 \\ 
 &&v_y=v_y^{(1)}(x+{W\over2}) -v_y^{(2)}({W\over2}-x) \nonumber\\
 &&a_0(x)=a_0^{(1)}(x+{W\over2})+a_0^{(2)}({W\over2}-x)-a_0^{(2)}(W) \nonumber 
\end{eqnarray}
where $\Theta$ is the step function and where we assumed $W\ll L$. 
The gauge fixing in the present case is achieved through the requirements that $a_0$ 
vanishes on the left wall and $a_y$ equals $-W/2$ there. The way $\theta$ is modified for the  
Coulomb interaction reflects the change in the electrochemical potential of one edge due to the 
electrostatic potential induced by the other. Inserting  expressions  (\ref{bsol1}) in the CSLG 
equations (\ref{bfield}) - (\ref{sch}) one finds that for the short range interaction 
they are satisfied up to exponentially small terms if $W\gg \alpha^{-1}$,  where $\alpha$  is the
decay constant  of the single edge  static  solution   defined  previously.  In  the 
Coulomb case  the set (\ref{bsol1}) is a  static  solution accurate up to terms of the order 
$\lambda_c |x_{1,2}|/W$  which we assume to be small .  The quantization of the Hall conductance
 is seen by integrating Eq. (\ref{efield}) with the result that the total current through the bar is 
$I=a_0(-{W\over2})-a_0({W\over2})=\mu_1-\mu_2-2\lambda_c(x_2-x_1)\ln L/W=
\tilde\mu_1-\tilde\mu_2+2\lambda_c(x_2-x_1)\ln W \equiv \nu(e^2/ {h})V_{\rm Hall}$
where in the last step we restored the units of dimensions. In the Coulomb case the logarithmic 
dependence on the width of the bar indicates that the current flows predominantly in the bulk.

We note that the sum and the difference of the values of $x_1$ and $x_2$ determine the charge of
and the current through the sample respectively. In particular $x_1=-x_2$ corresponds to the 
physically relevant neutral Hall bar.
For the set of symmetric solutions with $x_1=x_2$ the edge currents
balance each other while the asymmetric solutions $(x_1\neq x_2)$ represent a system through
 which a net current is flowing.

Turning  to the edge excitations  in the  Hall bar we can now  separately  use the derivatives of the 
static solution (\ref{bsol1}) with respect to either $x_1$ or $x_2$ to  attempt forming  gapless modes 
a la Eq. (\ref{lsol}) which will be concentrated {\em on one or another edge} of the bar  and 
propagating each in an opposite direction.  However  this prescription fails to satisfy the RPA equations
 in the case of the   Coulomb interaction.  Integrating the  linearized  continuity equation  in this 
case gives  the current in the $x$ direction (\ref{vxfield}) which does not vanish on the right edge. 
 We will now demonstrate that this
problem is cured if  the zero mode fields $\delta\rho^{(s)}$, $\delta v_y^{(s)}$  and $\delta a_0^{(s)}$
  in the ansatz (\ref{lsol})  are  taken as   linear combinations $\partial_{x_1}\rho + \beta 
\partial_{x_2}\rho$, etc., of the derivatives of the static solution (\ref{bsol1}). This  ansatz will 
now describe coupled density waves on opposite  edges propagating in the {\em same} 
direction. Substituting such  combinations   into Eq.~(\ref{lsch}) and assuming $k\ll W^{-1}$ we find
 to first  order in $k$  the  dispersion  relation which depends on the mixing coefficient $\beta$,     
\begin{equation}
\label{omcond}
{\omega\over{k}}=-{\partial\tilde\mu_1\over{\partial x_1}}+2\lambda_c\left[
(1+\beta)\ln\left({2e^{-\gamma}\over{|k|}}\right)-\beta \ln W\right] \; . 
\end{equation} The current   $\rho v_x$,  Eq. 
 (\ref{vxfield}) vanishes on the left wall by construction.  The value of $\beta$ is found by  
demanding that it will  also vanish  on the  right wall. This condition leads to a quadratic equation 
in $\beta$ giving the solutions  
\begin{equation}
\label{beta}
\beta_{\pm}=-Z\pm(Z^2-1)^{1\over 2} \;\;\; {\rm with} \;\;\; Z={\ln\left({2e^{-\gamma}\over{|k|}}\right)-
{1\over{4\lambda_c}}\left({\partial\tilde\mu_1\over{\partial x_1}}+{\partial\tilde\mu_2\over
{\partial x_2}}
\right)\over{\ln\left({2e^{-\gamma}\over{|k|W}}\right)}}\; ,
\end{equation}
and corresponding dispersion relations
\begin{equation}
\label{omcb}
\omega_{\pm}=-{1\over 2}\left({\partial\tilde\mu_1\over{\partial x_1}}-{\partial\tilde\mu_2\over{\partial x_2}}
\right)k\pm 2\lambda_c k\, \ln\left({2e^{-\gamma}\over{|k|W}}\right)
(Z^2-1)^{1\over 2}
\end{equation}
In the case of the short range interaction, $\lambda_c\rightarrow 0$, these solutions tend to
$\beta=0$ and $\beta=-\infty$ corresponding to modes concentrated on either the left or 
the right edge. The dispersion relations of these modes are the expected 
$\omega=-(\partial \mu_1/\partial x_1)k$ and $\omega=(\partial \mu_2/\partial x_2)k$. As anticipated 
the long range Coulomb forces result in inter-edge interaction which produces eigenmodes living 
simultaneously on both edges. This phenomenon has been predicted 
in Refs.\cite{Wen91,Oreg95} within the framework of the Luttinger model of the edge excitations 
in the QHE. It can be interpreted as a Coulomb drag of charges on one edge by the traveling density 
fluctuations on the other. Indeed $\beta_{\pm}$ are negative showing that the edges oscillate out of 
phase. The effect is 
the strongest for $k\rightarrow 0$ when $\beta_{\pm}$ tend to $-1\pm {\rm O}\left(|\ln k|^
{-{1\over 2}}\right)$. For a given mode the direction of 
propagation of the excitation is determined by the edge with the larger amplitude of charge 
fluctuations while the presence of the other edge reduces the velocity. For $k$  increasing 
towards $W^{-1}$ $\beta_{+}$ decreases in magnitude while $\beta_{-}=\beta_{+}^{-1}$ 
changes correspondingly in an opposite manner. Simultaneously the frequencies 
of the modes approach their single edge values indicating a decoupling 
of the edges in this limit.

For a neutral Hall bar sustaining a small current (i.e. small $x_1$) the current is given  
approximately by $I\simeq 2x_1(\partial\tilde\mu_1/\partial x_1)_{x_1=0}-4\lambda_c x_1\ln W$. 
In this region we find numerically that $\partial\tilde\mu_1/
\partial x_1\approx a+b\, x_1$ where $a\approx -1.5+0.75(\lambda_c+\lambda_c^2)$ and $b$ is a constant 
close to unity over the range of $\lambda_c$ presented in Fig. \ref{fig2}. 
These observations enable us to rewrite the dispersion relations (\ref{omcb}) 
in the following way
\begin{equation}
\label{iomcb}
\omega_{\pm}=\omega_c l\, k\left\{{b\pi\nu^{-1}\over{2\lambda_c\ln(W/l)-a}}\,
{I\over{e\omega_c}}\pm 2\lambda_c\sqrt{\left[\ln\left({2e^{-\gamma}\over{|k|l}}\right)-
{a\over{2\lambda_c}}\right]^2-\ln^2\left({2e^{-\gamma}\over{|k|W}}\right)}\right\}\; ,
\end{equation}
showing that when there is no current flowing in the bar the two modes 
travel in opposite directions with equal velocities. For a nonzero total current this symmetry is 
broken and one of the modes is carried with the flow while the other is retarded by it.
The dependence of the dispersion relation on the current suggests the possibility of a parametric 
excitation of these edge waves by driving an alternating current through the sample. 
For a FQHE sample with a length of 1mm the typical frequency of these modes is of the order 
of a few GHz. However by applying an additional constant current one can expect to excite the retarded 
branch at much lower frequencies. For the type of samples discussed here the values of the current for 
which the two terms in (\ref{iomcb}) approximately cancel each other lie in the $\mu$A region.

We would like to thank A. M. Finkel'stein and Y. Oreg for many useful discussions.

\begin{figure}
\caption{Density (solid lines) and current density (dashed lines) profiles for 
the case of short range interactions with $\lambda_s=1$. The highest curves 
correspond to $x_0=-2$. Consecutive curves differ by $\Delta x_0=1$. Note the reversal 
in the direction of the current when $x_0$ is increased. This behavior manifests a 
transition from a current originating from skipping orbits to a residual current due to  
uncompensated circular orbits on the rim of the Hall drop.}
\label{fig1}
\end{figure}

\begin{figure}
\caption{The chemical potential at the edge as a function of $x_0$ for short 
range interactions with $\lambda_s=1$ (solid line). For the  
Coulomb interaction we present $\tilde\mu=\mu+2\lambda_c x_0 \ln L$ 
in the realistic case  $\lambda_c=0.03$ (dashed line) and for comparison also 
the cases $\lambda_c=1$ (dashed-dotted line) and $\lambda_c=3$ (dotted line).}
\label{fig2}
\end{figure}

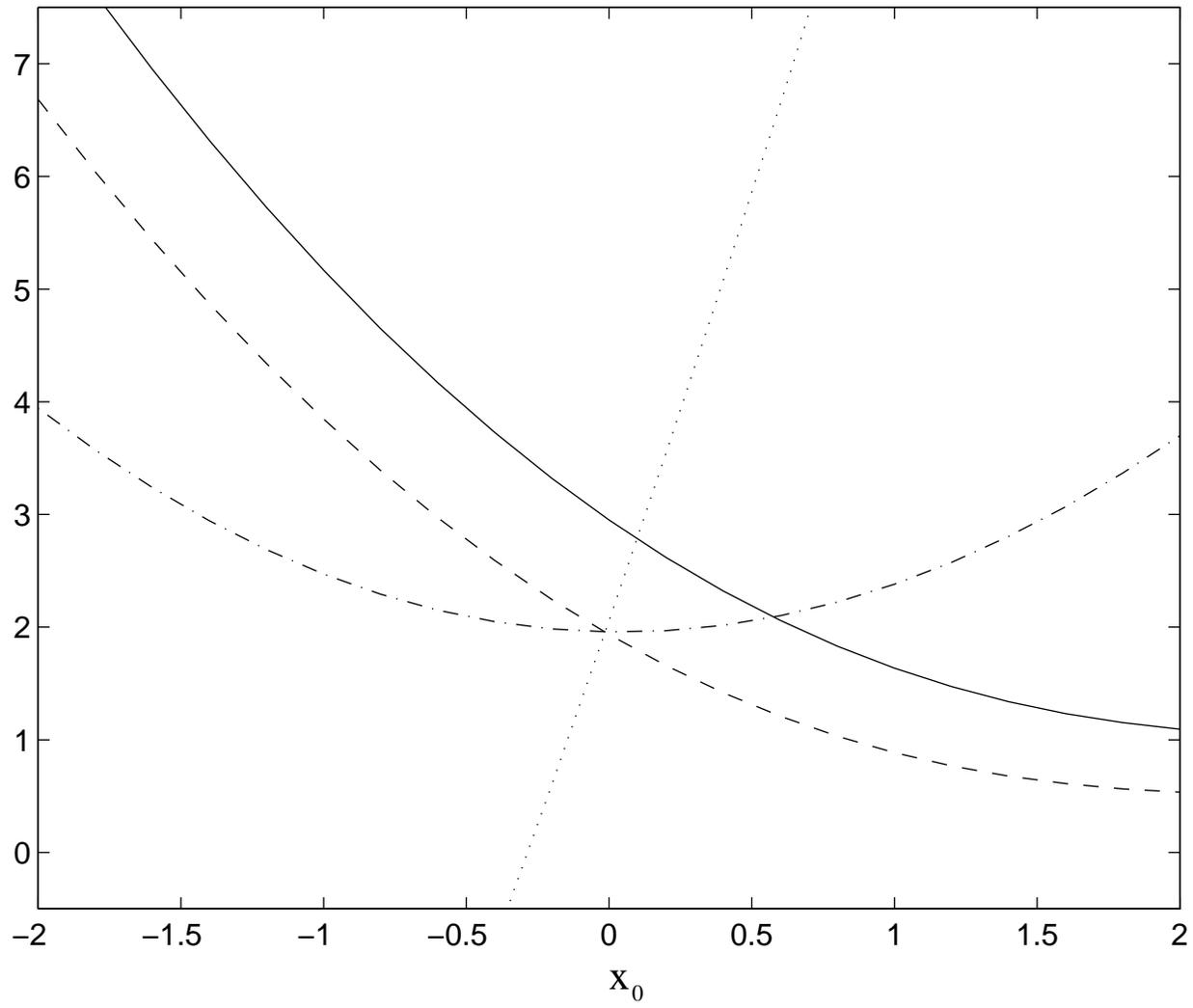
\end{document}